\documentclass[]{spie}  

\usepackage{amsmath,amsfonts,amssymb}
\usepackage{graphicx}
\usepackage[colorlinks=true, allcolors=blue]{hyperref}
\usepackage[mathscr]{eucal}
\usepackage{amsfonts}
\usepackage{mathtools}
\usepackage{xcolor}
\usepackage{floatrow}
\newcommand{\matr}[1]{\mathbf{#1}}
\renewcommand{\vec}[1]{\boldsymbol{#1}}

\title{Demonstrating predictive wavefront control with the Keck II near-infrared pyramid wavefront sensor}

\author[a]{Rebecca Jensen-Clem}
\author[b]{Charlotte Z. Bond}
\author[b]{Sylvain Cetre}
\author[a]{Eden McEwen}
\author[b]{Peter Wizinowich}
\author[b]{Sam Ragland}
\author[c]{Dimitri Mawet}
\author[a]{James Graham}
\affil[a]{Department of Astronomy, University of California at Berkeley, Berkeley, CA 94720}
\affil[b]{W. M. Keck Observatory, 65-1120 Mamalahoa Hwy., Kamuela, HI, USA 96743}
\affil[c]{Astronomy Department, California Institute of Technology, 1200 E. California Blvd.,
Pasadena, CA, USA 91125}

\authorinfo{Further author information: send correspondence to rjensenclem@berkeley.edu}

\begin{document} 
\maketitle

\begin{abstract}
The success of ground-based instruments for high contrast exoplanet imaging  depends on the degree to which adaptive optics (AO) systems can mitigate atmospheric turbulence. While modern AO systems typically suffer from millisecond time lags between wavefront measurement and control, predictive wavefront control (pWFC) is a means of compensating for those time lags using previous wavefront measurements, thereby improving the raw contrast in the post-coronagraphic science focal plane. A method of predictive control based on Empirical Orthogonal Functions (EOF) has previously been proposed and demonstrated on Subaru/SCExAO. In this paper we present initial tests of this method for application to the near-infrared pyramid wavefront sensor (PYWFS) recently installed in the Keck II AO system. We demonstrate the expected root-mean-square (RMS) wavefront error and contrast benefits of pWFC based on simulations, applying pWFC to on-sky telemetry data saved during commissioning of the PYWFS. We discuss how the performance varies as different temporal and spatial scales are included in the computation of the predictive filter. We further describe the implementation of EOF pWFC within the PYWFS dedicated real-time controller (RTC), and, via  daytime testing at the observatory, we demonstrate the performance of pWFC in real time when pre-computed phase screens are applied to the deformable mirror (DM). 

\end{abstract}


\section{INTRODUCTION}
\label{sec:intro}  
The last twenty years of astronomy have seen a revolution in planetary science, with over 3000 exoplanets discovered around nearby stars. This flood of new worlds includes planets unlike any found in our own Solar System, from Jupiter-mass planets located less than 0.1 AU from their suns\cite{1995Natur.378..355M}, to exotic rocky worlds twice as massive as the Earth\cite{2012Natur.482..195F}. While our understanding of exoplanet demographics -- how often exoplanets of different masses, radii, and orbital periods are found around different stellar types -- has leapt forward in recent years, fundamental questions remain. For example, what are the dominant planet formation pathways? How do planets acquire their atmospheres? What are the roles of clouds in shaping planets’ atmospheres?  

These questions can only be answered through the spectroscopic characterization of exoplanets with a range of masses and separations from their parent stars. While each method of exoplanet detection plays a complementary role in filling out the exoplanet mass--separation parameter space, only the methods of transits and direct imaging allow for the possibility of spectroscopic characterization. Exoplanet imaging has the advantages of versatility -- the planet's orbit need not be restricted to the small range of inclination angles necessary for a transit -- and precision -- high resolution spectroscopy of planetary atmospheres is possible when planet light is separated from star light. Resolving exoplanets very close to their host stars, however, requires the largest aperture telescopes ($>10$-m), and today is largely the domain of ground-based astronomy. 

Any Earthbound observer who looks up on a clear, windy night can see that our own atmosphere drastically distorts starlight before it reaches our eyes or telescopes. By compensating for atmospheric turbulence with AO, we can restore an image of a star to its proper diffraction-limited shape. Even so, light from possible planets will be obscured by the diffraction rings of the host star. A coronagraph is used to suppress that starlight, creating a dark hole in the final science image where a planet can be observed. So far, only extremely hot, massive worlds have been directly imaged, while lower mass objects, such as planets that may have formed via core accretion like those in our own Solar System, remain hidden in the glare of their host stars. 

Improvements in AO correction are required to directly image such planets. One limiting factor of a typical AO system is the temporal error, where the few milliseconds that elapse between sensing the wavefront and correcting it are enough for the wavefront to undergo significant evolution. Such time lags are disastrous for exoplanet imaging because they often cause light leakage into the regions closest to the star where planets are most likely to be found (e.g. the ``wind butterfly" effect\cite{1998SPIE.3353.1038R,2018SPIE10703E..6EM}). \emph{Predictive} WFC is any approach that seeks to compensate for these time lags by predicting the shape of the wavefront one or more time lags into the future based on past wavefront measurements. Since the widespread adoption of AO in the astronomical community, a range of methods for pWFC have been proposed. For example, Poyneer et al.~2007\cite{2007JOSAA..24.2645P}, 2008\cite{2008JOSAA..25.1486P} describe Fourier-based wind predictive control by identifying the windspeeds at different layers in the atmosphere using real-time AO telemetry data; Sivo et al.~2016\cite{2016SPIE.9909E..4YS} discuss the real-time implementation of a Linear-quadratic-Gaussian (LQG) tip-tilt controller for Gemini/GeMS; Correia et al.~2017\cite{Correia:17} present a distributed Kalman filter to reduce aniso-servo-lag and aliasing errors while simultaneously minimizing the total residual wavefront variance. Recently, Guyon \& Males (2017)\cite{2017arXiv170700570G} presented Empirical Orthogonal Functions, a method of linear control that is now in regular use at SCExAO\cite{2018SPIE10703E..1EG}. 

Here, we present the integration of EOF predictive control into the W. M. Keck Observatory PYWFS RTC\cite{2018SPIE10703E..39C} and the associated performance gains through on-sky telemetry data and daytime testing. Sections \ref{sec:EOF} and \ref{sec:kpic} briefly review EOF predictive control and the KPIC instrument, respectively; Section \ref{sec:telemetry} describes the improvements in RMS wavefront error (WFE) and contrast when EOF-pWFC is applied to on-sky telemetry data; and finally, Section \ref{sec:bench} describes the integration of EOF-pWFC into the Keck PYWFS control architecture and preliminary verification using the instrument's internal white light source.

\subsection{Predictive Control with Empirical Orthogonal Functions}
\label{sec:EOF}

The EOF approach is described in detail in Guyon \& Males (2017)\cite{2017arXiv170700570G} and is briefly reviewed here. A wavefront measurement at a given time $t$ is represented by a vector $\tilde{\vec{w}}(t)$ of length $m$, where $m$ is e.g. the number of actuators or Zernike modes used to represent the wavefront. Starting with the most recent wavefront measurement, we append the $n$ most recent wavefront measurements into a ``history vector" $\vec{h}(t)$:
\begin{equation}
\label{eqn:history}
\vec{h}(t) =\begin{bmatrix} \tilde{\vec{w}}_0(t) \\ \tilde{\vec{w}}_1(t) \\ \vdots \\ \tilde{\vec{w}}_{m-1}(t) \\ \tilde{\vec{w}}_0(t-dt) \\ \vdots \\ \tilde{\vec{w}}_{m-1}(t-dt) \\ \vdots \\ \tilde{\vec{w}}_{m-1}(t-(n-1)dt) \end{bmatrix}
\end{equation}
where $dt$ is the time step between wavefront measurements. We also define the AO system latency, or the time lag between measuring and correcting the wavefront, as $\delta t$. We will represent the wavefront one $\delta t$ into the future as the linear sum of past wavefronts. Hence, we seek to find the linear filter that minimizes the distance, in the least square sense, between a linear filter applied to the history vector and the wavefront one time lag into the future:
\begin{equation}
min_{\matr{F}^i} < || \matr{F}^i \vec{h}(t) - \vec{w}_i(t + \delta t) ||^2 >_t.
\end{equation}
We then find the filter using a ``training set" of $l$ previous history vectors and the wavefront measurements one time lag after each of those history vectors. These are the matrices $\matr{D}$ and $\tilde{\matr{P}}$, respectively. The filter solution is obtained from solving the least square problem:
\begin{equation}
min_{\matr{F}^i}  || \matr{D}^T {\matr{F}^i}^T - {\tilde{\matr{P}_i}}^T||^2
\end{equation}
with the filter solution:
\begin{equation}
\matr{F}^i = \left( (\matr{D}^T)^+ {\tilde{\matr{P}}_i}^T \right)^T
\end{equation}
where $(\matr{D}^T)^+$ is the pseudo-inverse of $\matr{D}$. While Guyon \& Males (2017)\cite{2017arXiv170700570G} compute the pseudo-inverse using singular value decomposition (SVD), we adopt a  regularized least-squares inversion  method:
\begin{equation}
\label{eqn:regls}
\matr{F}^i = \matr{P} \matr{D}^{T}(\matr{D} \matr{D}^{T} + \alpha \matr{I})^{-1}
\end{equation}
where $\matr{I}$ is the identity matrix and $\alpha$ is a constant. We find that this approach computes the filter 28 times faster than the SVD approach, as implemented in numpy\cite{numpy11} with $m=400$, $n=10$, and $l=60000$. 

\subsection{The Keck Planet Imager and Characterizer}
\label{sec:kpic}

Our test of pWFC relies on the  Keck Planet Imager and Characterizer (KPIC\cite{2018SPIE10703E..06M}), an on-going series of upgrades to the W.~M.~Keck II AO system aimed at exoplanet imaging and high resolution spectroscopy. As part of KPIC, the PYWFS was installed in the Keck II AO system in  September of 2018\cite{2018SPIE10703E..1ZB}, and underwent several successful commissioning runs in 2019~\cite{Bond19}. The implementation of pWFC described in this paper is made possible by the PYWFS RTC\cite{2018SPIE10703E..39C}. The RTC hardware is based on an x86 architecture linux server with a Geforce GTX 1080 ti Graphical Processing Unit; its software is based on the Compute and Control for Adaptive optics (CACAO) real-time control software package\cite{2018SPIE10703E..1EG}. To minimize impact on the existing AO system, the PYWFS RTC interfaces with the existing Keck MGAOS RTC rather than the DM and tip/tilt hardware directly. Section \ref{sec:bench} describes our implementation of EOF pWFC in the PYWFS RTC.


\section{APPLYING PREDICTIVE CONTROL TO KPIC PYWFS TELEMETRY}
\label{sec:telemetry}  

In order to demonstrate the benefits of pWFC to the Keck II PYWFS, we first apply predictive control ex post facto to on-sky closed loop telemetry data. Our dataset is a  time series of DM commands and residual wavefront measurements, both in units of voltage, taken during on-sky closed loop operation of the Keck II near-IR PYWFS on 2019-04-20. When reporting the RMS WFE, we convert from voltage to microns by the DM conversation factor $c=0.6 \mu$m$/$Volt. The dataset includes 120k time steps in total, as the system was run at $1\,$kHz for two minutes. At each time step, there are 349 DM commands and residual wavefront measurements, corresponding to the 349 actuators on the Keck DM. Because we are interested in predicting the next wavefront based on the state of the atmosphere only, we estimate the  open-loop wavefronts as:
\begin{equation}
\label{eqn:pseudo}
    \vec{w}(t) = -\mbox{DMC(t)} + \mbox{resid(t)}
\end{equation}
where DMC(t) is the DM command and resid(t) is the measured residual WFE at time t. Not captured in this framework are the modes that were unsensed or poorly sensed by the PYWFS. 

We now consider the effects of closed-loop integrator and predictive control on our incoming wavefronts, $\vec{w}(t)$. At each time step, we subtract a correction from the incoming wavefront; the result is the ``true" residual wavefront. To compute the ``measured" wavefront residual, we consider a PYWFS with four faces and a modulation amplitude of $3\frac{\lambda}{D}$. We model the phase reconstructed by such a PWFS using a simplified form of the Fourier-based PYWFS formalism described in Fauvarque et al.~(2017)\cite{Fauvarque17}; this is the ``measured" residual wavefront. For the case of non-predictive, integrator control, the correction is the sum of the previous time-step's correction and a gain factor times the measured residual (we choose a gain of $0.5$). To obtain the training set that will be used to compute the predictive filter, we save a series of ``pseudo open loop" wavefront measurements in closed-loop non-predictive control: the previous time step's correction plus the measured residual. Once the filter has been computed, we enter closed loop predictive control by assembling a history vector from the last $n$ pseudo open loop wavefronts, and take the correction to be the product of the filter and the history vector (i.~e.~gain$\,=1$ for the predictive case). We note that we are considering a ``zonal" approach, as the history vectors are populated by actuator voltages.

The choices of the training set size $l$, filter order $n$, and degrees of freedom $m$ are described below. The regularization parameter $\alpha$ is set to one for all cases described in this section. All residuals reported in this section are the true residuals rather than the measured residuals. 

\subsection{Exploring the Temporal Filter Parameters}

We first consider a prediction of all 349 actuator positions one time lag into the future given the past positions of all 349 actuators (i.e.~$m=349$). The peril of this approach is that the positions of actuators that are well separated from each other in the pupil plane may not be correlated, so including them in the same filter computation may serve only to introduce additional noise. The potential benefit of this approach is that all possible spatial correlations between actuators are taken into account when computing the filter. 

\begin{figure} [h]
\begin{center}
\includegraphics[width=\textwidth]{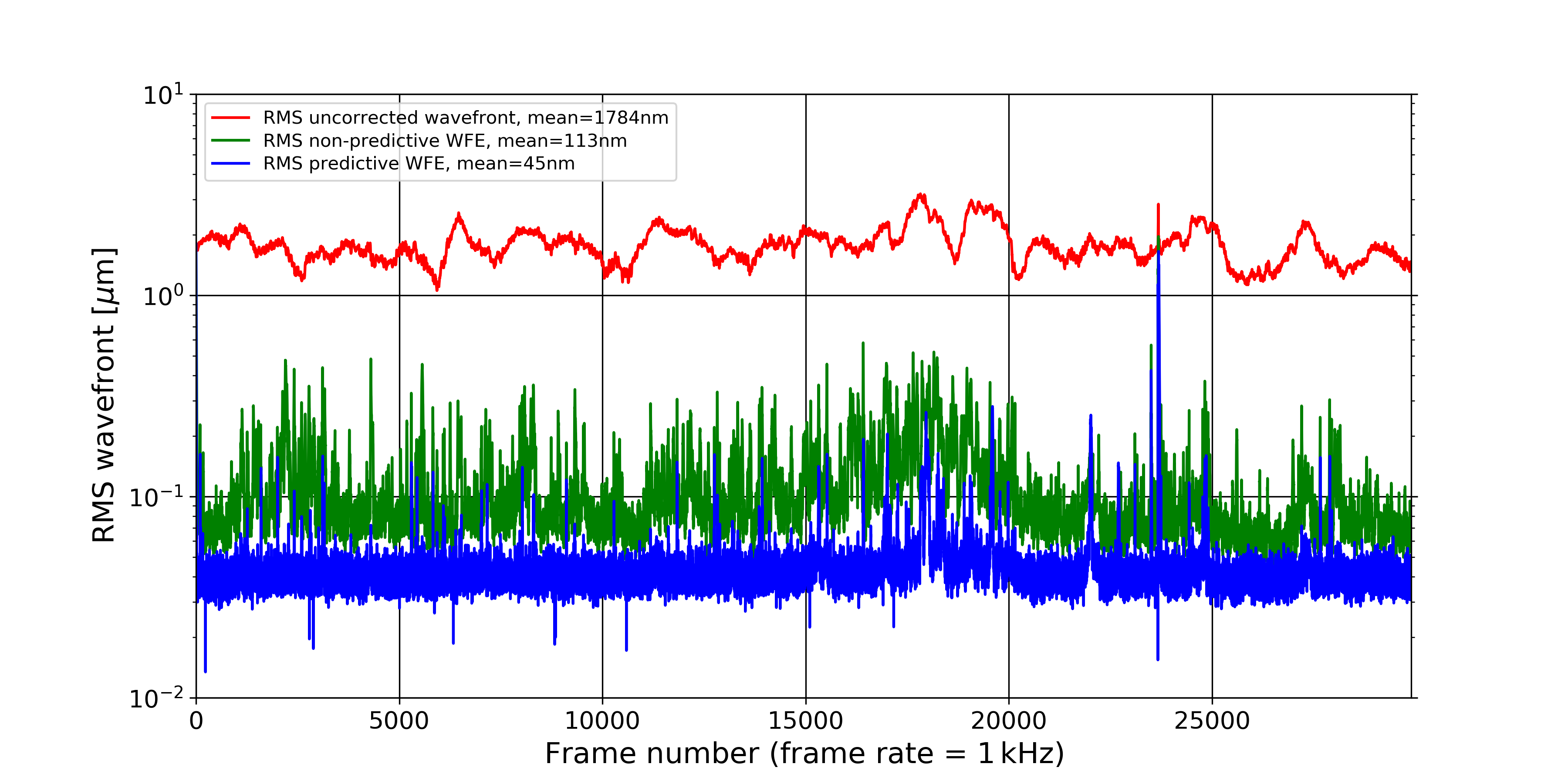}
\end{center}
\caption[] {\label{fig:telem} The RMS of the uncorrected wavefront given by Equation \ref{eqn:pseudo} (red), the integrator (non-predictive) residual (green), and the residual after applying the predictive filter (blue). Here, the training set length was $l=60000$, the filter order was $n=5$, and the degrees of freedom was $m=349$ DM actuators.}
\end{figure}

Following Section \ref{sec:EOF}, we construct a filter from the first 60 seconds of our telemetry data such that $m=349$, $n=5$, and $l=60000$, and apply the filter to the next 30 seconds, or 30k frames, of our telemetry data. Figure \ref{fig:telem} shows a per-frame comparison of the RMS error of the uncorrected wavefronts, the residuals after applying non-predictive control, and the residuals after applying predictive control. We note that the absolute scaling of the y-axis of Figure \ref{fig:telem} underestimates the total WFE that we would see on sky due to the limitations of our telemetry data set (described at the beginning of Section \ref{sec:telemetry}).


\begin{figure} [h]
\begin{center}
\begin{tabular}{cc}
\includegraphics[width=0.5\textwidth]{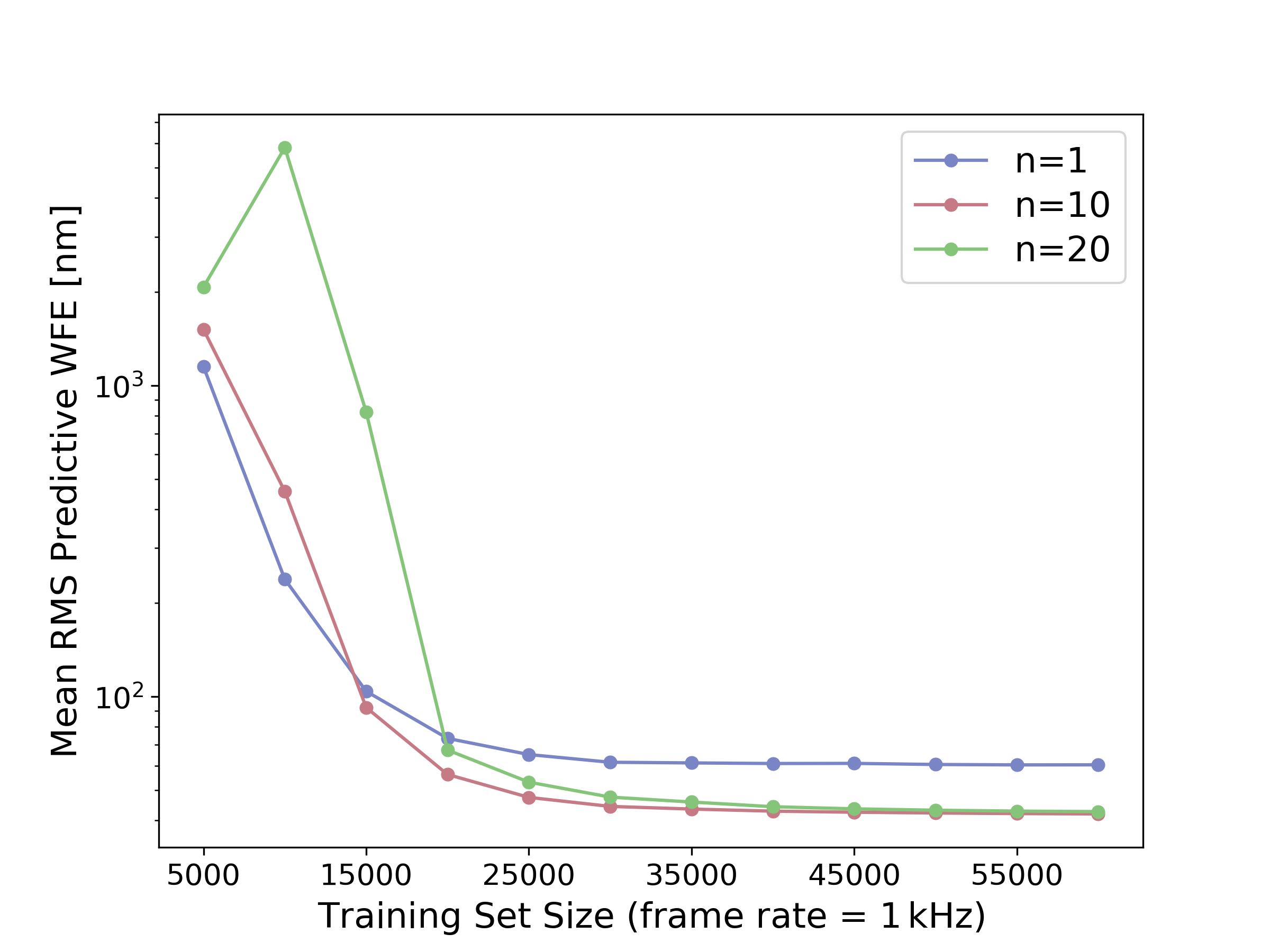} &
\includegraphics[width=0.5\textwidth]{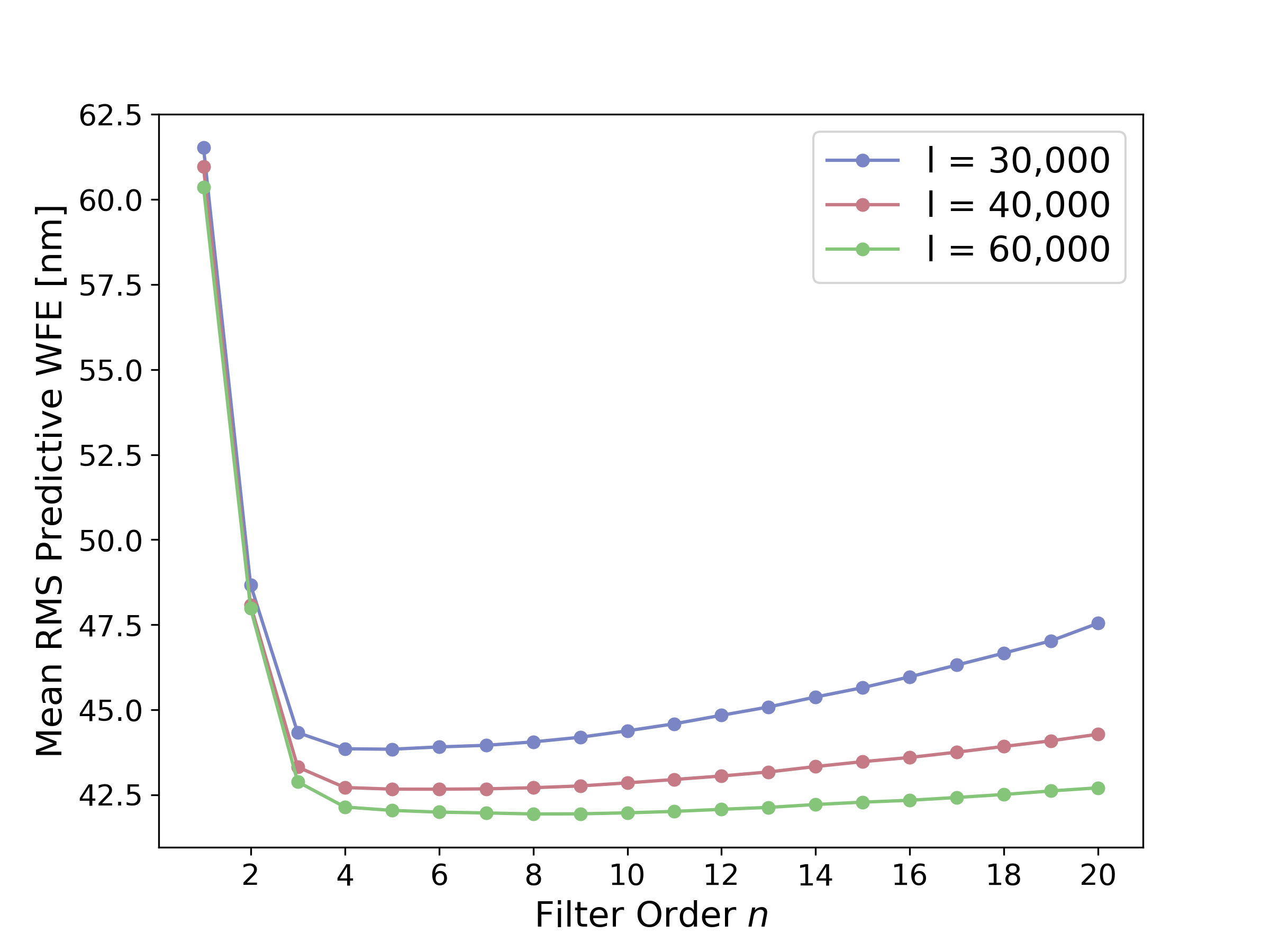} \\
(a) & (b) \\
\end{tabular}
\end{center}
\caption[] {\label{fig:varyln} (a) The mean residual WFE after the application of a predictive filter as a function of the training set length, $l$, and for three values of the filter order, $n$. (b) The mean residual WFE after the application of a predictive filter as a function of the filter order, $n$, and for three values of the training set length, $l$.}
\end{figure}

In Figure \ref{fig:varyln}a, we explore the performance of predictive filters constructed using different training set lengths, $l$, when applied to the same set of 30k test wavefronts. Each point in Figures \ref{fig:varyln}a is frame-by-frame wavefront RMS, averaged over the test set. We see that the performance of the filter levels off for $l>40000$. We further see that the filter performance suffers when the training set is too small (e.~g.~for $l<15000$ here).

The three lines in Figures \ref{fig:varyln}a show three different values of the filter order, $n$. Choosing a filter order that is too large degrades the performance of the filter due to the inclusion of wavefronts that are so far into the past as to be less correlated with the wavefront one time step into the future. However, including only the most recent wavefront ($n=1$) does not take advantage of the temporal correlations that enhance the performance of the $n=5$ filter.

These effects are more readily apparent in Figure \ref{fig:varyln}b, where we plot the mean WFE as a function of the filter order for three different training set lengths. The performance of the filter quickly improves as $n$ is increased from one to about $n=5$, and then gradually worsens as less correlated wavefronts from the more distant past begin to add noise. 

\subsection{Exploring the Spatial Filter Parameters}
In the previous section, we included all 349 DM actuators in the computation of our filter (i.e.~$m=349$), and varied the temporal parameters $n$ and $l$. We found that including too many time steps in the history vector degraded the performance of the filter. Here, we consider whether including too many or too few spatial elements in the history vector might also affect the filter's performance. 

\begin{figure} [h]
\begin{center}
\begin{tabular}{cc}
\includegraphics[width=0.5\textwidth]{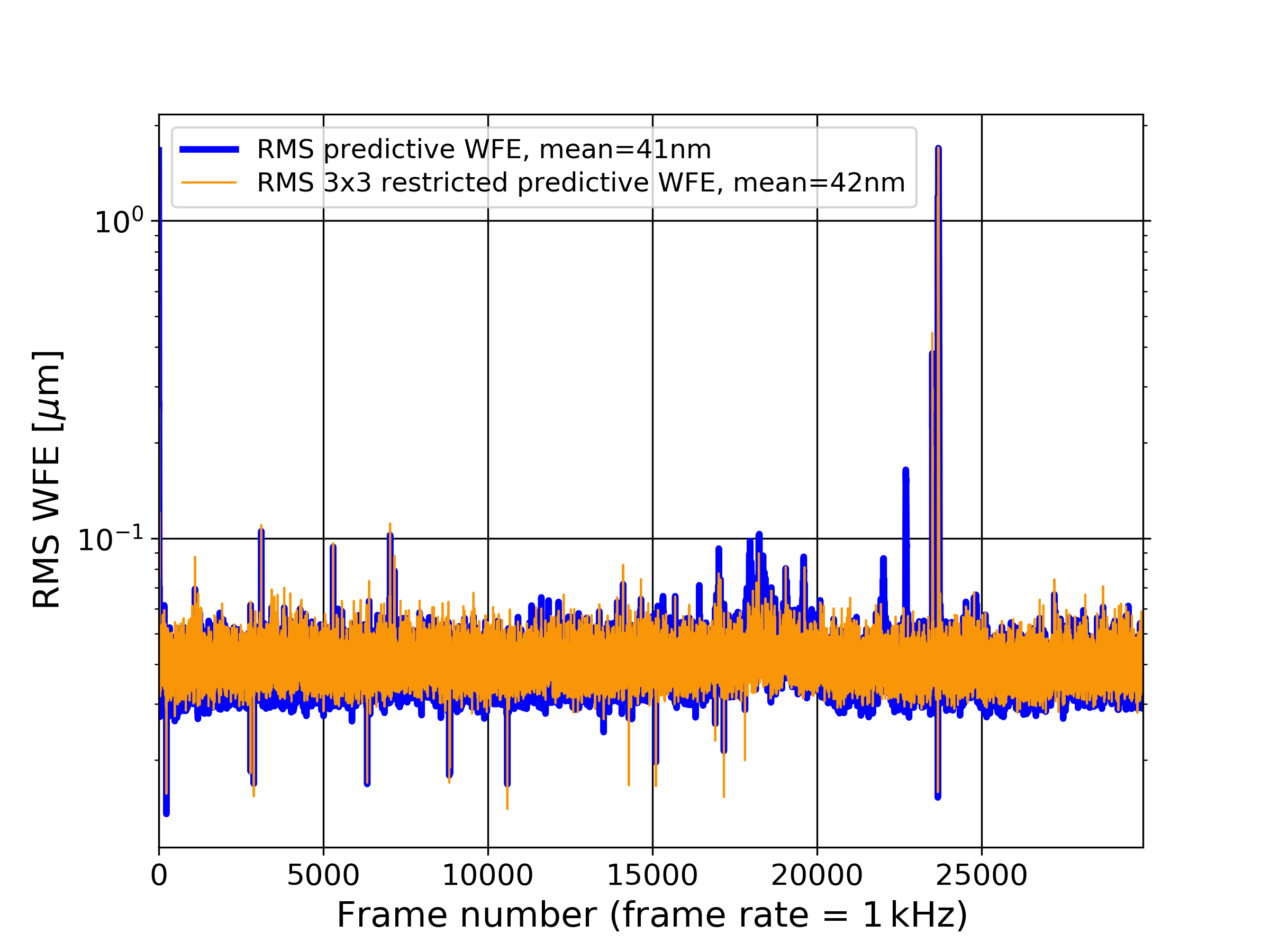} &
\includegraphics[width=0.5\textwidth]{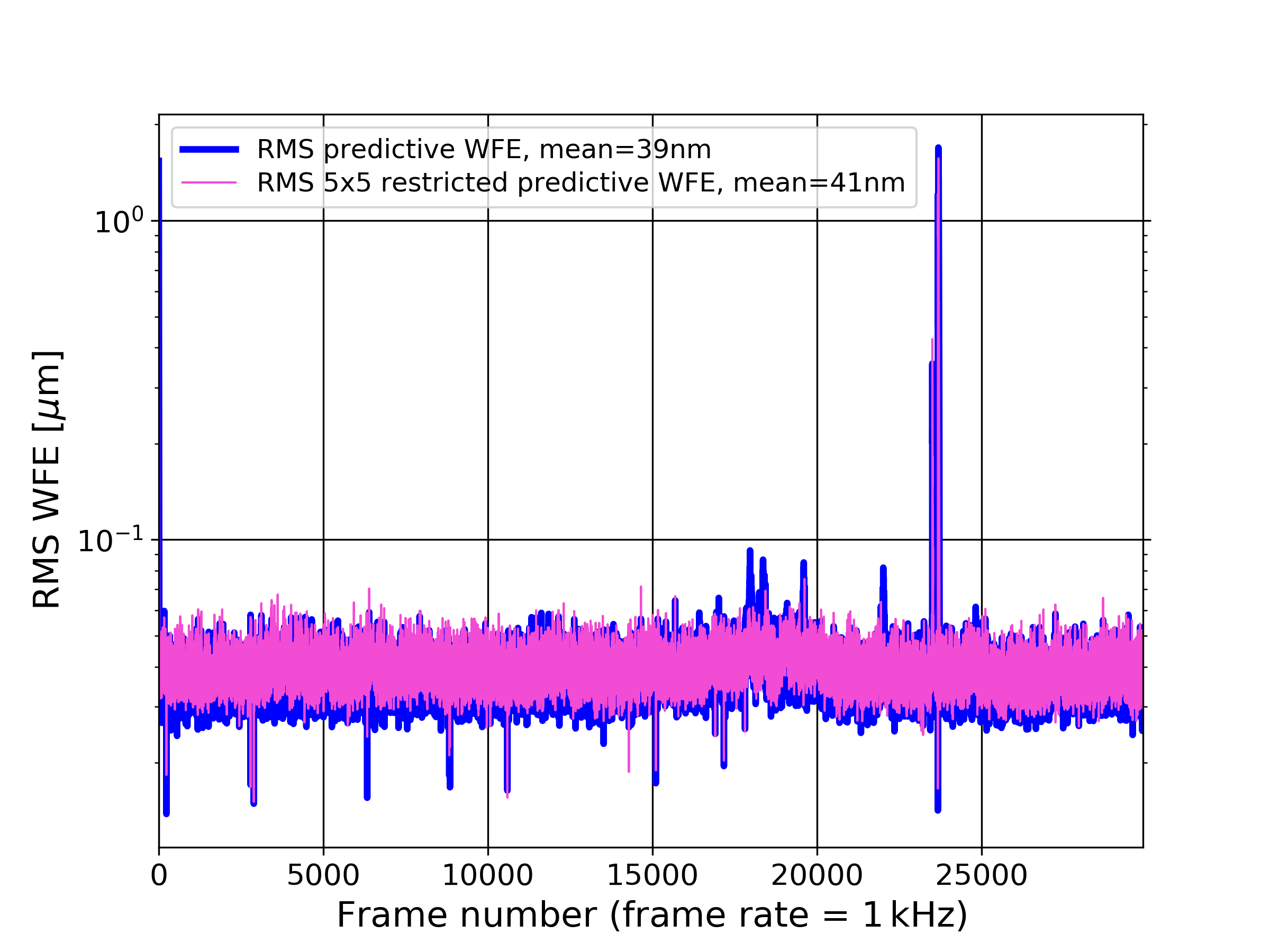} \\
(a) & (b) \\
\end{tabular}
\end{center}
\caption[] {\label{fig:segs} A comparison between the $m=349$, $m=9$ (left), and $m=25$ (right) predictive residuals ($n=5$ and $l=60$k for all cases shown here). The blue predictive residuals ($m=349$) are identical to those shown in Figure \ref{fig:telem}, with the exception that the number of actuators over which the per-frame RMS WFE was calculated was windowed to match the spatially restricted cases. }
\end{figure}

Instead of creating a single filter whose inputs and outputs include all 349 actuator positions, we now create a different filter for each individual actuator. The history vector now contains the actuator positions from either a $3\times3$ or a $5\times5$ box centered on the actuator of interest (see also the approach described in van Kooten et al.~2019\cite{vanKooten:19}). For simplicity, we disregard edge actuators where a complete box is not possible. We choose $n=5$ and $l=60000$ for all filters. We then compute the predictive residual for each actuator using its own filter, and compute the RMS of all such predictive residuals for a given time step. The results are shown in Figure \ref{fig:segs}, where the blue line is the same predictive residual based on the $m=349$ filter plotted in Figure \ref{fig:telem}, but with the same edge actuators discarded as in the $3\times3$ or $5\times5$ box cases. 

Figure \ref{fig:segs} shows that restricting the filter to $3\times3$ or $5\times5$ boxes centered on each actuator does not significantly improve the performance of the filter. The large RMS jumps are primarily due to the on-sky measured residuals (hence the open loop measurements in the history vector were contaminated). We do see improved performance in the spatially restricted filters at the location of the small hump in the residuals near time step $18$k.  The residual excursions to smaller values are present in all predictive residuals, and their origins are not clear. 

\subsection{Synthetic Contrast Curve}
\begin{figure} [h]
\begin{center}
\begin{tabular}{cc}
\includegraphics[width=0.35\textwidth]{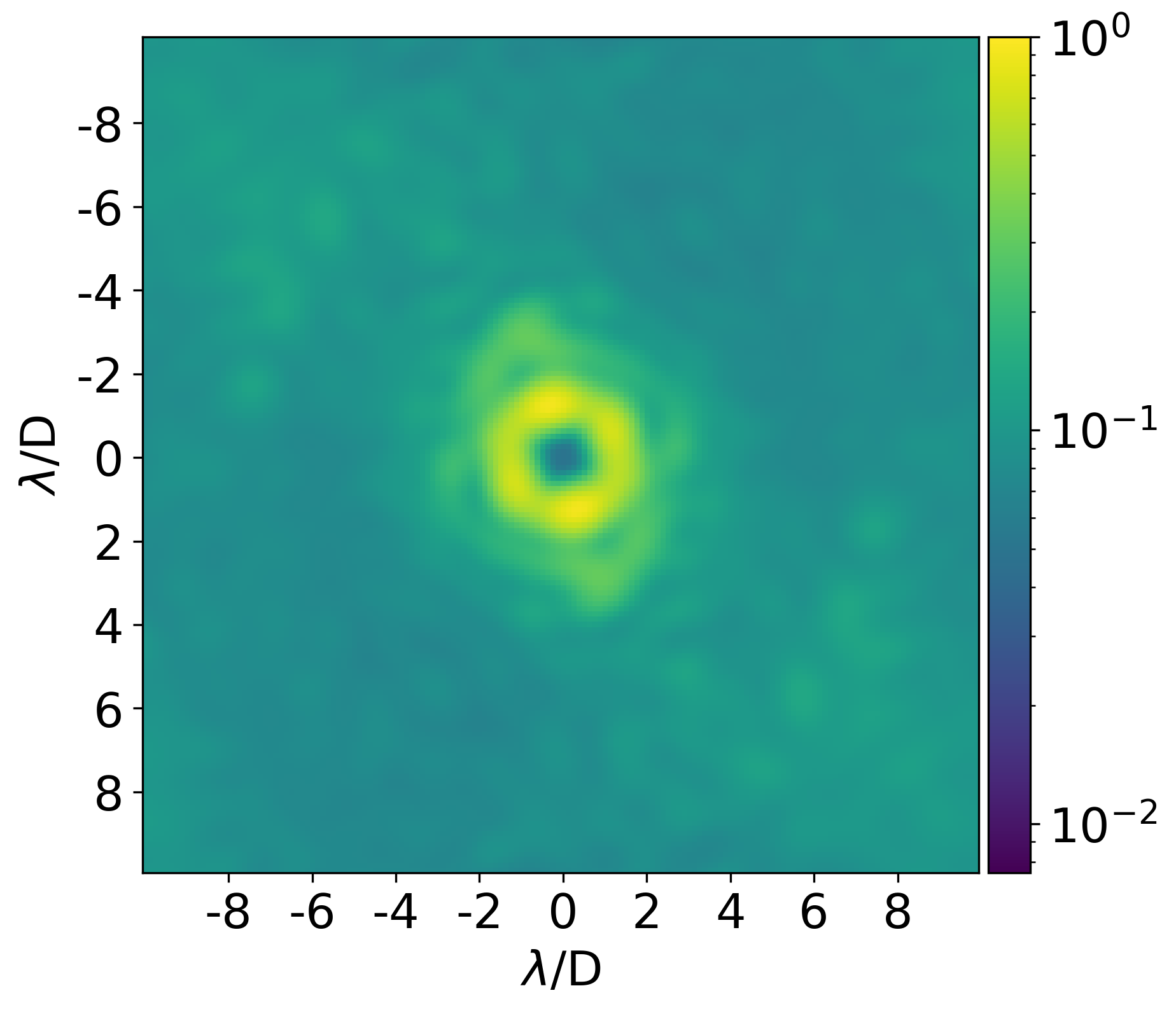} &
\includegraphics[width=0.35\textwidth]{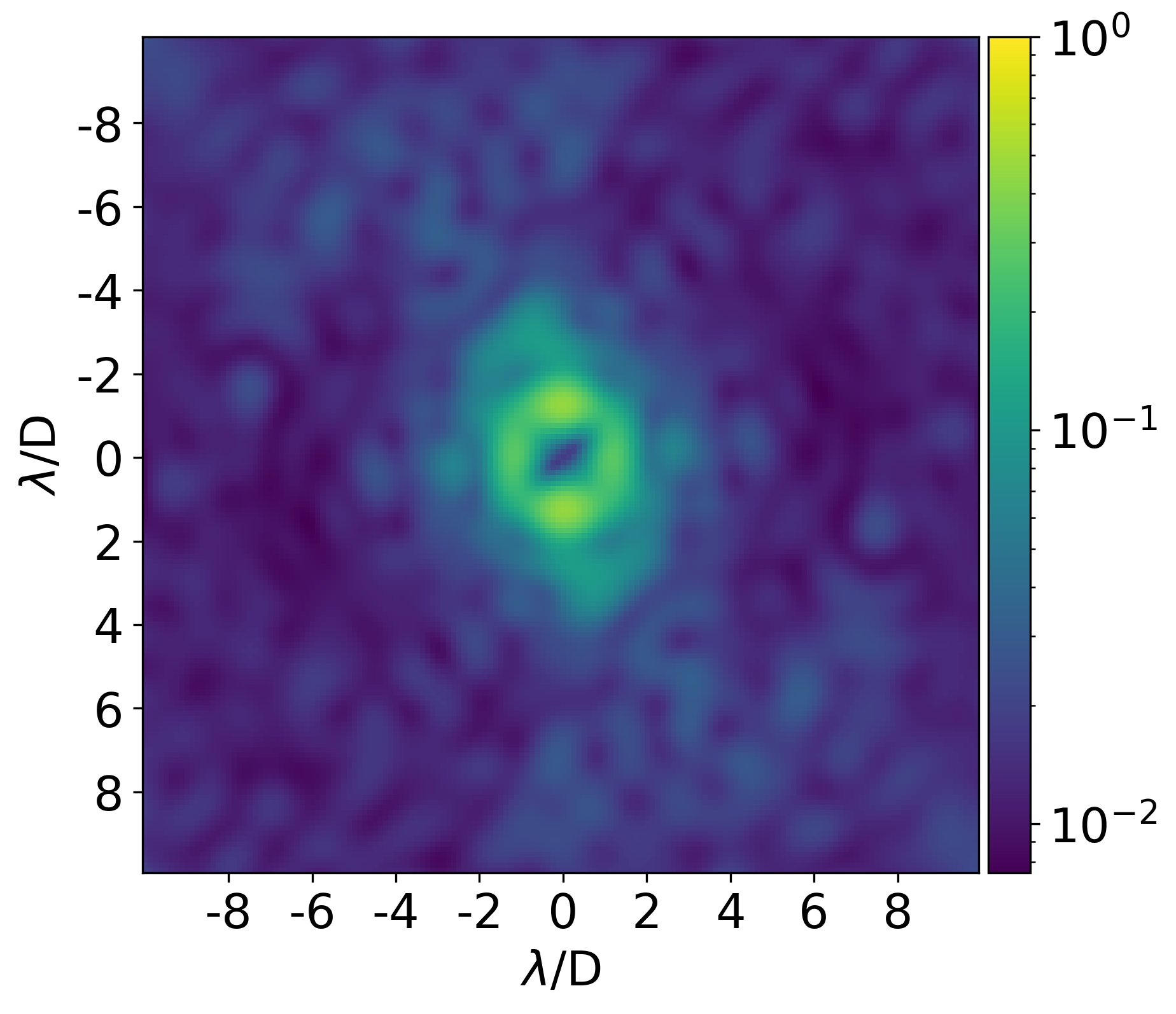} \\
(a)  & (b) \\
\end{tabular}
\end{center}
\caption[] {\label{fig:synthetic_psfs} Synthetic post-coronagraphic PSFs each representing a 30 second observation with non-predictive control (a) and predictive control (b).   }
\end{figure}

\begin{figure} [h]
\begin{center}
\includegraphics[width=0.75\textwidth]{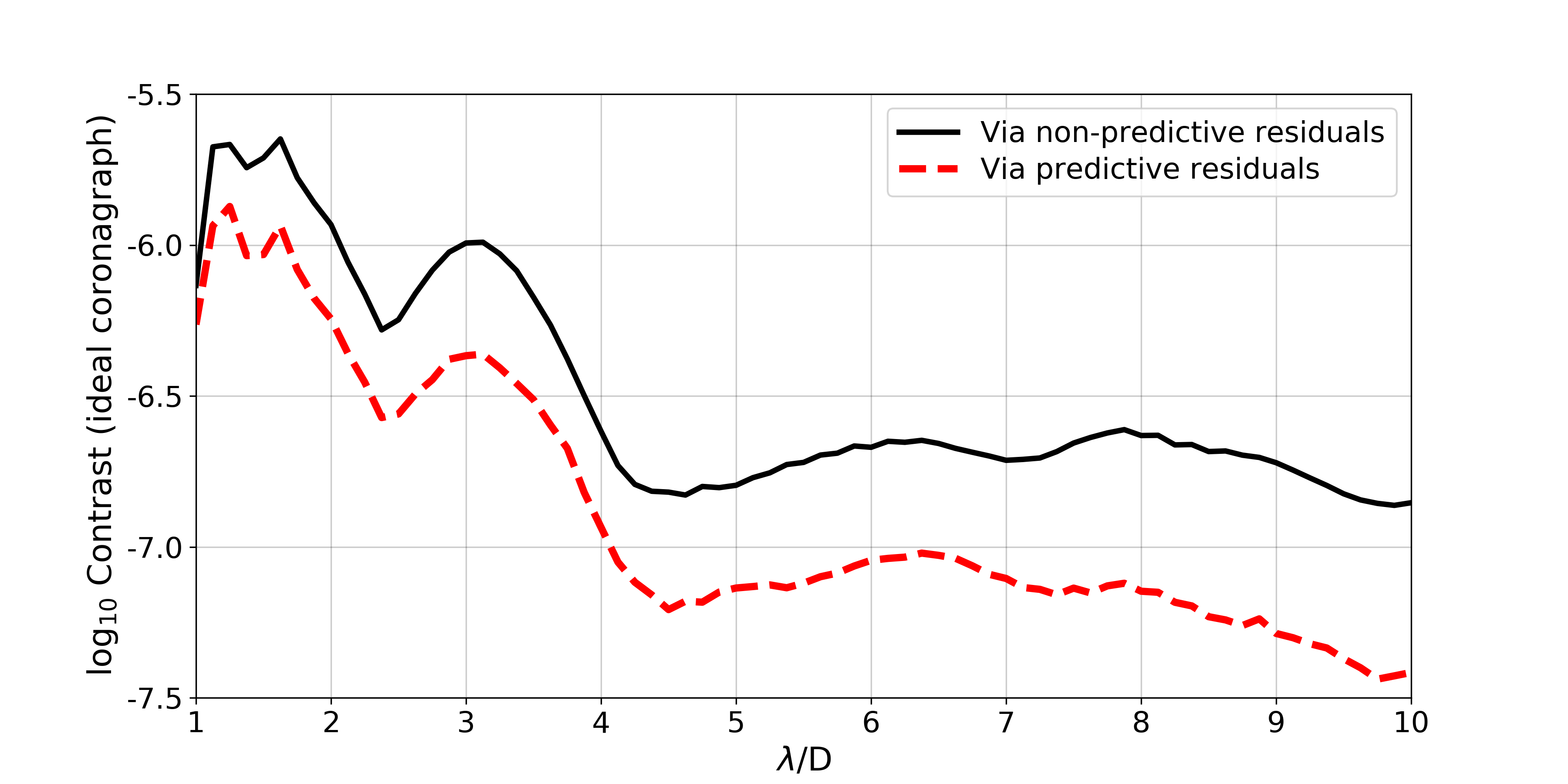}
\end{center}
\caption[]{\label{fig:contrast} A synthetic contrast curve at $\lambda=2.11\mu$m based on the non-predictive residuals and the residuals after applying the predictive filter ($m=349$, $n=10$, $l=60000$). }
\end{figure}

Given the improvement in RMS WFE due to pWFC shown in Figure \ref{fig:telem}, we now consider how these WFE improvements translate into contrast ratio improvements in the focal plane. To simulate contrast curves, we consider 30k electric fields with uniform amplitudes and phases equal to the non-predictive and predictive residual wavefronts represented in Figure \ref{fig:telem}. We then Fraunhofer propagate these electric fields to the focal plane and compute the intensity using hcipy\cite{por2018hcipy}. We create an ideal point spread function (PSF) by Fraunhofer propagating a clear aperture matching the shape of the valid DM actuators. We simulate the effect of an ideal coronagraph by subtracting this ideal PSF from the non-predictive and predictive PSFs. The summed resulting PSFs are shown in Figure \ref{fig:synthetic_psfs}. We then divide these sets of 30k PSF-subtracted focal plane images into two chunks of 15k images each, and sum each chunk to create two ``long exposure" focal plane images. We subtract the first long exposure from the second long exposure to simulate the effects of PSF calibration\cite{2017arXiv170700570G}. We then use VIP\cite{2017AJ....154....7G} to generate the contrast curves based on these final PSFs (Figure \ref{fig:contrast}). The contrast curve generated with predictive control represents about a $2.6$X improvement over the non-predictive contrast curve. 

Because we have idealized the effects of the coronagraph and PSF calibration, the contrast curves in Figure \ref{fig:contrast} should be interpreted relative to one another; the absolute scaling of the y-axis is not relevant here. Further, Figure \ref{fig:contrast} shows the contrast benefit due to pWFC when the non-predictive contrast curve is limited by time-delay errors only. Effects such as fitting error, non-common path aberrations, etc are not represented here. 

\section{IMPLEMENTING REAL-TIME PREDICTIVE CONTROL ON KPIC'S PYWFS}
\label{sec:bench}  

The new real-time controller (RTC) that supports the Keck II PYWFS is described in detail in Cetre et al.~(2018)\cite{2018SPIE10703E..39C} and reviewed briefly here. To avoid moving the DM cabling and to support easy switching between the PYWFS and the facility SHWFS, the PYWFS RTC interfaces with the observatory's existing wavefront controller (MGAOS\cite{Johansson:06}, developed by Microgate) rather than sending commands to the DM and tip/tilt mirror directly. The PYWFS RTC is based on an easily scalable x86 architecture linux server, and currently includes one Geforce GTX 1080 ti Graphical Processing Unit (GPU). The software is based on Compute And Control for Adaptive Optics (CACAO\cite{2018SPIE10703E..1EG}), an open source software package written in C and originally designed to support SCExAO. 

One of the challenges associated with the implementation of predictive control in the PYWFS RTC is to handle the command history (or ``history vector" as described by Equation \ref{eqn:history}) without impacting the RTC latency. For this reason we have chosen to store the history ring buffer directly in the GPU memory. The ``down time" of the loop (i.e.~the time between sending the DM commands and receiving the next set of pixels from the wavefront sensing detector) has been used to store the wavefront history in the ring buffer. Several new GPU routines, or ``kernels," were developed to support predictive control: (1) to download and organize the wavefront information in the GPU, and (2) to apply the predictive control via a matrix vector multiplication. Additional memory has been reserved in the GPU to store the filter matrix, and an additional interface to the RTC has been created to enable the predictive control. When predictive control is enabled, this implementation adds only $30\mu$s to the overall system latency, due to the matrix vector multiplication needed to apply the predictive filter. When in use, the kernels described above are executed in the normal AO correction sequence, just after the existing reconstruction step. 

\begin{figure} [ht]
\begin{center}
\includegraphics[width=\textwidth]{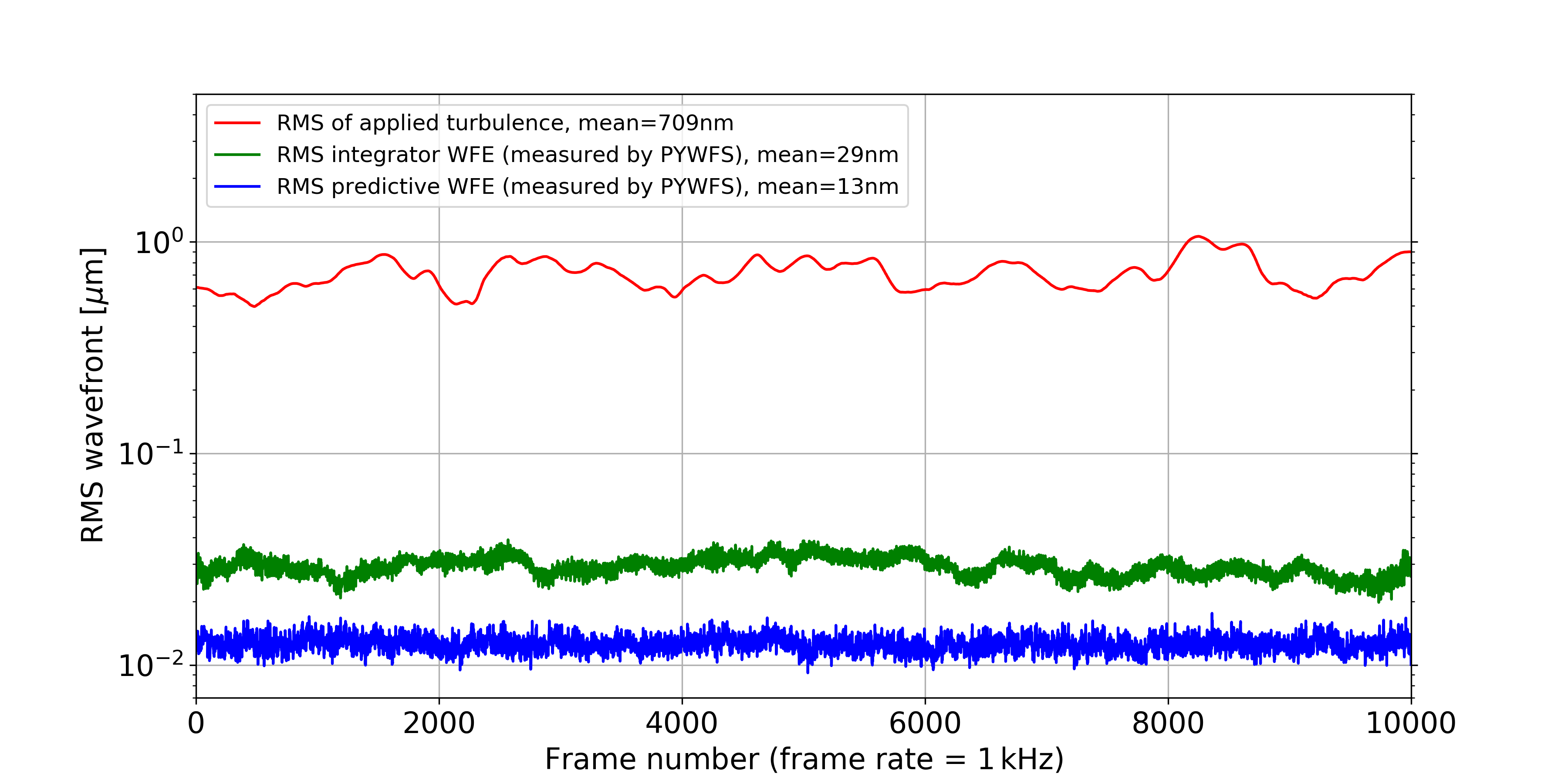}
\end{center}
\caption[] {\label{fig:benchdemo} The RMS of the turbulence screens applied to the Keck II DM (red), the RMS of the residual WFE as measured by the Keck II PYWFS when non-predictive control was enabled (green), and the RMS of the residual WFE as measured by the Keck II PYWFS when predictive control was enabled (blue).}
\end{figure}

We tested this implementation during the day using the Keck II AO system's internal white light source and by applying pre-computed phase screens to the DM at $1$kHz. These phase screens were generated using hcipy\cite{por2018hcipy}, and represent a single layer of turbulence following a Kolmogorov spectrum, with $r_0=16\,$cm, $L_0=20\,$m, and a wind velocity of $5\,$m$/$s. During one minute of closed-loop integrator control with $g=0.5$, we recorded the pseudo-open loop turbulence (Equation \ref{eqn:pseudo}) and computed the predictive filter based on these data ($l=60$k, $n=10$, $m=349$, $\alpha=2000$). We then enabled predictive control and recorded the residual wavefront error as measured by the PYWFS for $10\,$s. The results are shown in Figure \ref{fig:benchdemo}: the time-averaged predictive residual is about $2.2\times$ smaller than the time-averaged non-predictive residual. We note that the PYWFS modal gains are set up such that the absolute values of the RMS WFE in Figure \ref{fig:benchdemo} are underestimated; because the non-predictive and predictive cases are affected by the modal gains in the same way, however, their RMS WFE values can be evaluated relative to one another.

\section{FUTURE WORK}
\label{sec:future}  

Improving our contrast ratios at the smallest inner working angles is critical for directly imaging new classes of exoplanets with current $10$-m telescopes and future $30$-m telescopes. In this paper, we've described a preliminary implementation of EOF pWFC for the Keck II near-IR PYWFS and demonstrated its promise for improving the system's residual WFE using (1) ex post facto analysis of on-sky telemetry data, and (2) real time predictive control of the AO system during day-time operations, in conjunction with the system's internal white light source and phase screens corresponding to single-layer frozen flow turbulence applied to the DM. 

We plan to extend our day-time demonstrations of predictive control to include phase screens corresponding to multi-layer frozen flow turbulence as well as the uncorrected wavefronts estimated from on-sky telemetry data (Equation \ref{eqn:pseudo}). During these extended tests, we will also investigate the effects of ``mixing" the predictive and non-predictive commands to improve the stability of the correction (e.g.~an approach in which $10\%$ of the correction comes from the linear controller and $90\%$ comes from the predictive solution), as well as varying the matrix regularization parameter $\alpha$ given in Equation \ref{eqn:regls}. 

In the Fall of 2019, we plan to demonstrate pWFC on sky over a range of atmospheric conditions, stellar magnitudes, and zenith angles. We will correlate the spatial and temporal filter parameters discussed in this paper with atmospheric coherence lengths and coherence times. Finally, we will measure the gain in raw contrast due to pWFC using the K- and L-band vortex coronagraphs and NIRC2 detector. 


\acknowledgments     
 
The predictive wavefront control demonstration is funded by the Heising-Simons Foundation. The authors would like to thank Marcos van Dam (Flat Wavefronts) for his suggestion to use the regularized least-squares inversion method to compute the predictive filter. The W. M. Keck Observatory is operated as a scientific partnership among the California Institute of Technology, the University of California, and the National Aeronautics and Space Administration. The Observatory was made possible by the generous financial support of the W. M. Keck Foundation. The near-infrared pyramid wavefront sensor is supported by the National Science Foundation under Grant No. AST-1611623. The PYWS camera was provided by Don Hall as part of his National Science Foundation funding under Grant No. AST 1106391. Support for R.~J.~-C.~was provided by the Miller Institute for Basic Research in the Sciences. This work benefited from the NASA Nexus for Exoplanet System Science (NExSS) research coordination network sponsored by the NASA Science Mission Directorate. The authors wish to recognize and acknowledge the very significant cultural role and reverence that the summit of Maunakea has always had within the indigenous Hawaiian community. We are most fortunate to have the opportunity to conduct observations from this mountain. 

\bibliography{report} 
\bibliographystyle{spiebib}

\end{document}